\documentclass[12pt]{iopart}
\usepackage{amssymb}
\usepackage{graphicx}
\begin{document}


\title[Solving relativistic hydrodynamic equation with magnetic field]{Solving 
relativistic hydrodynamic equation in presence 
of magnetic field for phase transition in a neutron star}

\author{Ritam Mallick$^1$, Rajesh Gopal$^2$, Sanjay k Ghosh$^2$, 
Sibaji Raha$^2$ \& Suparna Roychowdhury$^3$}
\address{$1$ Institute of Physics, Bhubaneswar 751005, Orissa, INDIA}
\address{$2$ Centre for Astroparticle Physics and
Space Science; Bose Institute; Block - EN, Sector V; Salt Lake; Kolkata -
700091; INDIA}
\address{$3$ Department of Physics; St. Xaviers College; 
Kolkata - 700016; INDIA} 

\ead{ritam.mallick5@gmail.com}

\begin{abstract}
Hadronic to quark matter phase transition may occur 
inside neutron stars (NS) having central densities of the order of 3-10 times 
normal nuclear matter saturation density ($n_0$). The transition 
is expected to be a two-step process; transition from hadronic to 2-flavour matter and two-flavour to $\beta$ equilibrated charge neutral three-flavour 
matter. In this paper we concentrate
on the first step process and solve the relativistic
hydrodynamic equations for the conversion front in presence of high magnetic field. Lorentz force due 
to magnetic field is included in the energy momentum tensor by averaging over the polar angles. We 
find that for an initial dipole configuration of the magnetic field 
with a sufficiently high value at the surface, velocity of 
the front increases considerably.
\end{abstract}



\maketitle

\section{Introduction}
Study of strongly interacting matter at high density and/or temperature is
of immense current interest. As proposed in \cite{witten,haesel} a form of matter 
consisting of almost equal numbers of up (u), down (d) and strange (s) 
quarks, may be the true ground state of strongly interacting matter at high densities and/or temperature.
This conjecture is supported by Bag model calculations
\cite{farhi} for certain range of values of the strange quark mass
and the strong coupling constant. This form of bulk
matter consisting of u, d, and s quarks is referred to as \char`\"{}strange quark 
matter (SQM)\char`\"{}.

The extreme conditions of high density and/or temperature can be envisaged through at least two scenarios;
one being the terrestrial lab in which such conditions can be reproduced (like
controlled study of elementary particle interactions at high energies) 
and the other being the astrophysical lab of compact objects. Some 
advancement has already been made to recreate the deconfinement 
transition in the 
collision of heavy nuclei in high energy accelerators {\it{e.g.}} at CERN-SPS 
in Geneva and RHIC in Brookhaven national laboratory (BNL). 
Along with the above terrestrial laboratories, compact stars having 
central density as high as 3-10 times nuclear matter saturation density,
can be used to test nature of strongly interacting matter.   

High density inside NSs is expected to induce a transition to quark matter. If 
SQM is the true ground state of strongly interacting matter, as 
conjectured by Witten \cite{witten}, then there is a novel possibility
that some NSs can get completely converted to SS.
Some tentative SS candidates are the compact 
objects associated with X-ray bursters GRO $J1744-28$ \cite{cheng}, 
SAX $J1808.4-3658$ \cite{chin} and X-ray pulsars Her $X-1$ \cite{dey}.

A mechanism of the phase transition from NS to SS was first proposed by 
Alcock et al. \cite{alcock}. 
They argued that the conversion
process starts when the star comes in contact with an external strange 
quark nugget seed. Glendenning (\cite{glen}) on the other hand suggested
that the conversion may begin when there is a sudden spin down of 
a neutron star. The abrupt change in the angular velocity of the star
suddenly increases the central (or core) density thereby
triggering the phase transition process.  

Several authors has studied the hadronic to quark matter conversion. 
Olinto \cite{olinto} considered the phase 
transition of hadronic to SQM as a weak interaction process and evaluated 
the front velocity assuming it to be a slow-combustion. 
Olesen and Madsen \cite{madsen} and Heiselberg  et al. 
\cite{byam} found the speed of conversion to be in the 
range between 10 m/s to 100 km/s. 
Conversion of hadronic to two-flavour quark matter in the astrophysical scenario was studied by 
Collins and Perry \cite{collins}. They argued that the hadronic matter, at first, undergoes deconfinement to quark matter and then eventually decays to three-flavour
matter through weak process. A non-relativistic hydrodynamic stability 
analysis of such combustion process was done by  
Horvath and Benvenuto \cite{ben2}. Cho  et al. \cite{cho} improved it 
further and used a relativistic framework to examine the
conservation condition of energy-momentum and baryonic density flux
across the conversion front. 
Tokareva  et al. \cite{toka} studied the conversion process assuming it to be 
detonation shock front. On the other hand, 
Berezhiani et al. \cite{berez}, Bombaci  et al. \cite{bomb} 
and Drago  et al. \cite{drago}, related it with gamma ray bursts,
suggesting that the formation of SQM
may be delayed if the deconfinement process takes place through a
first order transition \cite{alam}. The two-step process
was again recently considered by Bombaci et al. and other 
authors \cite{bomb2,mintz}. More recent development on the subject includes the
dynamical simulations of the combustion process both in 1D (\cite{nieb}) and
in 3D (\cite{herzog}).

Most of the works however have not incorporated some of the important 
features of NS, namely the rotational and the general relativistic 
(GR) effects. We have tried to incorporate these effects systematically in a 
series of works. We have shown \cite{ritam1} that the 
conversion process is most likely a two-step process. 
Following Glendenning we heuristically assumed the existence of conversion 
front originating at the center and propagating outwards. 
The first step involved the conversion of NS to a 
two-flavour QS. We assumed the central density to be high, ($3-10$) 
times normal matter density, and that the transition front proceeds as a
detonation. We found that within a few milliseconds the NS gets converted to a 
metastable two-flavour QS. The second step corresponds to 
the conversion of two-flavour quark matter to the stable three-flavour SQM, 
through weak interaction. We have also exhibited \cite{ritam2} the major 
role played by the GR effects in the rotating stars; the conversion 
fronts, corresponding to the nuclear to two-flavour matter conversion, 
propagate with different radial velocities in different directions. 

Identifying pulsars with rotating NS leads to the conclusion that NS have very 
high magnetic fields ($10^{10}-10^{12}$ G) at the surface. The origin of such 
magnetic fields in NS remains unclear till date; even if all magnetic lines 
of force were to be trapped during the collapse, one would estimate 
\cite{michel} a surface field of $10^8-10^9$ G. The observed slow-down rate 
of pulsars require surface magnetic fields in the range of $10^{10}-10^{12}$ G.
The canonical picture of the classical pulsar mechanism involves \cite{michel} 
a magnetic dipole at the center of a rotating NS. In one of our ongoing 
project \cite{ritam3} we have tried to 
include the effect of magnetic field in the general 
relativistic framework through the 
introduction of a Lorentz force on the equation of motion by hand.
As opposed to the present case where only dipolar configuration has been
assumed, in \cite{ritam3} we have studied other different magnetic field 
configuration as well. For the dipolar field the nature of front propagation 
remains more or less same (only slight change in velocity). 
One of the interesting findings of \cite{ritam3} 
is that if the magnetic field is of planar radial nature which decreases
towards the surface, the term due to the magnetic field opposes the front
velocity. If the magnetic field is quite high then it may stall the front 
inside the star and much higher values of magnetic field stops the front
from taking off.

In the present work our effort is to incorporate the effect of the magnetic field on the equation of
motion starting from the conservation equation. Moreover, we try to have an understanding of the angular dependence of the magnetic field 
by averaging the Lorentz force over the polar angles.
Since this is our first attempt, we have limited our study to the 
special relativistic case and present the findings after the incorporation of the 
magnetic field in the relativistic hydrodynamic equations for the 
conversion of nuclear matter to two-flavour quark matter. We have performed our calculation with zero   
viscosity and resistivity, that is in the ideal hydrodynamic approximation.

To study the conversion process,
we have considered the relativistic EOSs describing the forms of the matter
in respective phases. Due to the propagation of the shock wave, the
properties of the medium on either side of the shock exhibit a discontinuity.
We have studied the various conservation conditions on either side of the 
front. Development
of the conversion front, as it propagates radially through the model
star in the presence of magnetic field, has been examined. Here, we have considered 
very slowly rotating neutron star so that we can use the static limits.
In section II, we discuss the EOSs and the initial jump conditions.
In section III, we present the governing equations for the 
combustion front propagation by incorporating the effect of the Lorentz 
force due to the magnetic field. In section IV, we present the results 
obtained by solving the relativistic MHD 
equations. Conclusions that may be
drawn from these results are presented in the final
section.

\section{Initial jump condition}

The nuclear matter phase is described by 
the non-linear Walecka model \cite{walecka} and the quark phase 
by the MIT Bag model \cite{alcock}, quark numbers being fixed by the number of baryons
in the nuclear phase. We treat both 
the nuclear and quark phases as ideal fluids.

The nuclear matter EOS has been evaluated using the nonlinear Walecka
model \cite{walecka}. The Lagrangian density in this model is given
by: 

\[
\mathcal{L}(x)=\sum_{i}\bar{\psi_{i}}(i\gamma^{\mu}\partial_{\mu}-m_{i}
+g_{\sigma i}\sigma+g_{\omega i}\omega_{\mu}\gamma^{\mu}
-g_{\rho i}\rho_{\mu}^{a}\gamma^{\mu}T_{a})\psi_{i}-
\frac{1}{4}\omega^{\mu\nu}\omega_{\mu\nu}\]

\[
+\frac{1}{2}m_{\omega}^{2}\omega_{\mu}\omega^{\mu}
+\frac{1}{2}(\partial_{\mu}\sigma\partial^{\mu}\sigma
-m_{\sigma}^{2}\sigma^{2})-\frac{1}{4}\rho_{\mu\nu}^{a}\rho_{a}^{\mu\nu}
+\frac{1}{2}m_{\rho}^{2}\rho_{\mu}^{a}\rho_{a}^{\mu}\]

\begin{equation}
-\frac{1}{3}bm_{n}(g_{\sigma N} {\sigma})^{3}-\frac{1}{4}C(g_{\sigma N} {\sigma})^{4}
+\bar{\psi_{e}}(i\gamma^{\mu}\partial_{\mu}-m_{e})\psi_{e}\qquad\qquad
\label{1}\end{equation}

The Lagrangian in equation (1) includes nucleons (neutrons and protons),
electrons, isoscalar scalar, isoscalar vector and isovector vector
mesons denoted by $\psi_{i}$, $\psi_{e}$, $\sigma$, $\omega^{\mu}$
and $\rho^{a,\mu}$, respectively. The Lagrangian also includes cubic
and quartic self interaction terms of the $\sigma$ field. The parameters
of the nonlinear Walecka model are meson-baryon coupling constants,
meson masses and the coefficient of the cubic and quartic self interaction
of the $\sigma$ mesons (b and c, respectively). The meson fields interact
with the baryons through linear coupling. The $\omega$ and $\rho$
meson masses have been chosen to be their physical masses. The rest
of the parameters, namely, nucleon-meson coupling constant 
($\frac{g_{\sigma}}{m_{\sigma}},\frac{g_{\omega}}{m_{\omega}}$
and $\frac{g_{\rho}}{m_{\rho}}$) and the coefficient
of cubic and quartic terms of the $\sigma$ meson self interaction
(b and c, respectively) are determined by fitting the nuclear matter
saturation properties, namely, the binding energy/nucleon (-16 MeV), baryon density
($\rho_{0}$=0.17 $fm^{-3}$), symmetry energy coefficient (32.5 MeV),
Landau mass (0.83 $m_{n}$) and nuclear matter incompressibility (300 MeV).

\begin{figure}
\vskip 0.2in
\centering
\includegraphics[width=3.0in]{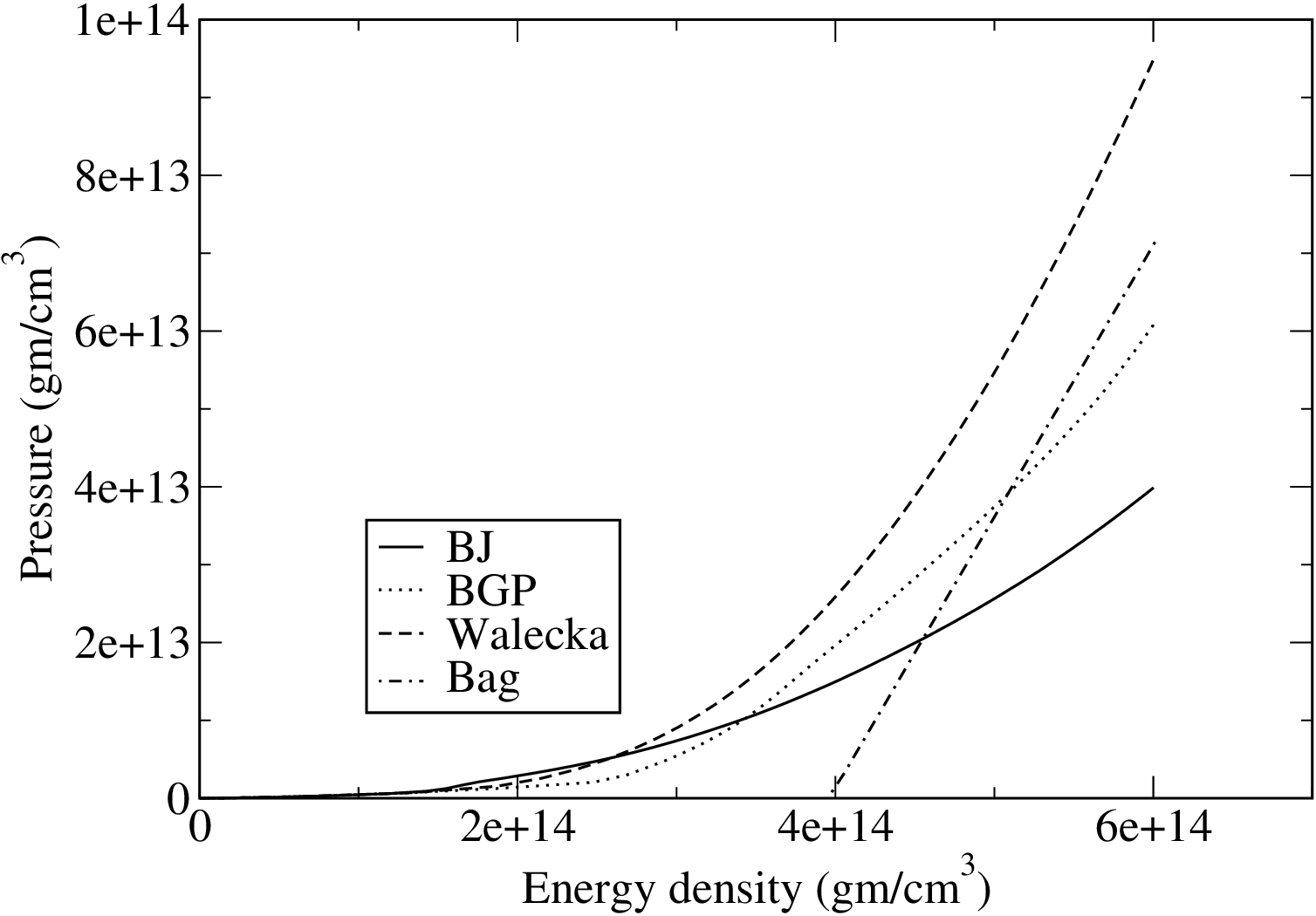}
\caption{Pressure vs. energy density plot for different model EOS.}
\label{eos}
\end{figure}

In the present paper, we first consider the conversion of nuclear
matter, consisting of only nucleons (\textit{i.e.} without hyperons)
to a two-flavour quark matter. The final composition of the quark matter
is determined
from the nuclear matter EOS by enforcing the baryon number conservation
during the conversion process.
That is, for every neutron two down and one up quarks and for every
proton two up and one down quarks are produced, electron number being
same in the two phases. The mass of the up and down quarks are taken to 
be $5$ MeV and $10$ MeV respectively. The value of the bag constant is taken to $160$ Mev.
The pressure-density relation of the two models are depicted in fig. \ref{eos}. 

Recently, after the discovery  
of high-mass pulsar PSR J1614-2230 \cite{demorest} 
the EOSs describing the interior of a compact star have been put to severe
constraint. Mostly EOS with 'exotic' matter
such as hyperons and kaon condensates which cannot generate high-mass stars
within a range of central densities are now under speculations. 
In \cite{demorest} typical values of the central density of J1614-2230, for the
allowed EOSs is in the range 2$n_0$ - 5$n_0$. On the other hand, consideration 
of the EOS independent analysis of \cite{lattimer2005} sets the upper limit of 
central density at 10$n_0$. 

\begin{figure}
\vskip 0.2in
\centering
\includegraphics[width=3.0in]{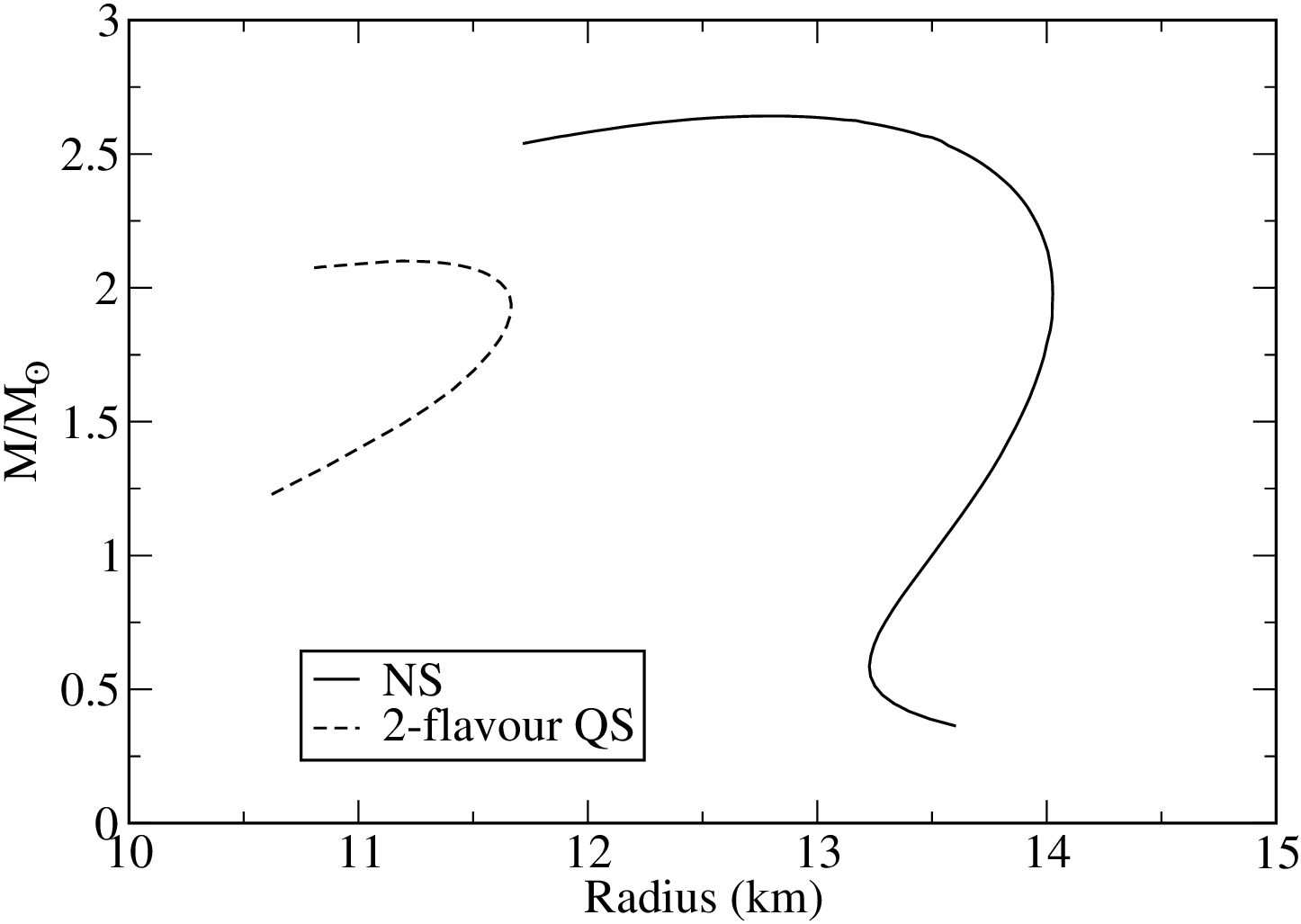}
\caption{Mass-radius plot for NS and two-flavour quark star, with bag constant
$160$ MeV.}
\label{mass-rad}
\end{figure}

We have shown the mass-radius curve in fig. \ref{mass-rad} for both hadronic 
and metastable two-flavour quark matter EOS. 
In Table \ref{table} we have given the
number density, mass and radius of stars made of 
hadronic matter, two-flavour metastable 
as well as three-flavour, $\beta$ equilibrated charge neutral matter. 
It is found that though both the hadronic as well as two-flavour matter 
with bag pressure of $160$ MeV are within the range of parameters as 
suggested by Demorest et al., three-flavour matter with the same bag 
pressure has smaller mass. A lower
bag pressure of $145$ MeV is needed to nearly reach the limits of 
\cite{demorest}.
Here we should mention that similar calculation done by other authors 
agrees with our results given in the Table \ref{table}. 
Recent calculation by Zudnik et al., Bauswein et al. and 
Weissenborn et al. \cite{zdunik,bauswein,weissenborn} found that the 
maximum masses of the strange star is just
below 2 solar masses with bag pressure $145$ MeV. There they have used 
MIT bag model to construct the star with strange quark mass of $100$ MeV. In 
Table \ref{table} we have shown strange star masses of bag pressure $145$ 
MeV and $160$ MeV. We have shown results for strange quark masses 
of $50$ MeV, $100$ MeV, $150$ MeV and $200$ MeV, and our results are at 
per with the other previous calculations.
As the mass of strange quark mass is still debated, for a range of strange 
quark mass we find that the maximum mass of the strange star with bag pressure
$160$ MeV is $1.6$ times mass of sun ($M_{\odot}$), which is
smaller than that prescribed by 
Demorest et al.. The limitation of bag model itself, as has been shown in 
\cite{kurkela}, is another important issue. Under the present circumstances, 
one may have to consider other options such as a strongly interacting system, 
a more involved picture as given in \cite{kurkela} or other effective models 
such as PNJL \cite{tamal} to repeat the similar calculations. However, in 
the present work we have concentrated on the effect of magnetic field on the 
front using the simple bag model picture and the EOS here 
is needed for estimating the initial velocity of the front. The effect of 
bag pressure and hence equation of state is of course important as the 
physical values of velocity is obtained only within a narrow range of 
$\pm$5 MeV around $160$ MeV bag pressure \cite{ritam1}. We have repeated our 
calculation for various densities and discussed the results for 4$n_0$ for 
illustration. 
 
Following the picture of Glendenning \cite{glen} we have assumed that the phase
transition is initiated by a sudden spin down of the star.
So let us now consider the physical situation where a combustion front
has been generated in the core of the neutron star. This front propagates
outwards through the neutron star with a certain hydrodynamic velocity, 
leaving behind a u-d-e matter. We denote all the physical
quantities in the hadronic sector by subscript 1 and those in the
quark sector by subscript 2.

\begin{table}[t]
\begin{center}
\begin{tabular}[t]{|l|l|l|l|}
\hline
\multicolumn{4}{|c|}{Table} \\
\hline
EOS & $n_B/n_0$ & $M/M_{\odot}$ & $R$ in km \\ \hline
Walecka & $4$ & $1.9$ & $14$ \\
 & $6$ & $2.45$ & $13.7$ \\
 & $8$ & $2.58$ & $13.4$ \\ \hline
Bag 160 MeV 2-flav. & $4$ & $1.63$ & $11.3$ \\
 & $6$ & $1.99$ & $11.7$ \\
 & $8$ & $2.1$ & $11.6$ \\ \hline
Bag 145 MeV 2-flav. & $4$ & $2.07$ & $12.2$ \\
 & $6$ & $2.26$ & $12.0$ \\
 & $8$ & $2.32$ & $11.8$ \\ \hline
Bag 160 MeV 3-flav.$m_s$=50 MeV & $4$ & $1.05$ & $9.74$ \\
 & $6$ & $1.46$ & $9.95$ \\
 & $8$ & $1.59$ & $9.74$ \\ \hline
Bag 160 MeV 3-flav.$m_s$=100 MeV & $4$ & $0.92$ & $9.15$ \\
 & $6$ & $1.37$ & $9.51$ \\
 & $8$ & $1.51$ & $9.4$ \\ \hline
Bag 160 MeV 3-flav.$m_s$=150 MeV & $4$ & $0.88$ & $9.32$ \\
 & $6$ & $1.19$ & $9.59$ \\
 & $8$ & $1.32$ & $9.60$ \\ \hline
Bag 160 MeV 3-flav.$m_s$=200 MeV & $4$ & $0.79$ & $9.03$ \\
 & $6$ & $1.11$ & $9.34$ \\
 & $8$ & $1.23$ & $9.36$ \\ \hline
Bag 145 MeV 3-flav.$m_s$=50 MeV & $4$ & $1.75$ & $11.9$ \\
 & $6$ & $1.93$ & $11.7$ \\
 & $8$ & $1.97$ & $11.5$ \\ \hline
Bag 145 MeV 3-flav.$m_s$=100 MeV & $4$ & $1.67$ & $11.7$ \\
 & $6$ & $1.87$ & $11.4$ \\
 & $8$ & $1.92$ & $11.1$ \\ \hline
Bag 145 MeV 3-flav.$m_s$=150 MeV & $4$ & $1.64$ & $11.5$ \\
 & $6$ & $1.8$ & $11.4$ \\
 & $8$ & $1.86$ & $11.4$ \\ \hline
Bag 145 MeV 3-flav.$m_s$=200 MeV & $4$ & $1.54$ & $11.3$ \\
 & $6$ & $1.76$ & $11.2$ \\
 & $8$ & $1.81$ & $11.1$ \\ \hline
\end{tabular}
\caption{Tabulated ratio of number density, mass and radius for stars with
different EOS}
\label{table}
\end{center}
\end{table}

Quantities on opposite sides of the front
are related through the energy density, the momentum density and the
baryon number density flux conservation \cite{toka,landau2,gleeson}. 
In the rest frame of the combustion front, the velocities of the matter 
in the two phases, are written as \cite{landau2}:

\begin{equation}
v_{1}^{2}=\frac{(p_{2}-p_{1})(\epsilon_{2}+p_{1})}{(\epsilon_{2}
-\epsilon_{1})(\epsilon_{1}+p_{2})},
\label{5}\end{equation}

and \begin{equation}
v_{2}^{2}=\frac{(p_{2}-p_{1})(\epsilon_{1}+p_{2})}{(\epsilon_{2}
-\epsilon_{1})(\epsilon_{2}+p_{1})}.
\label{6}\end{equation}

\section{Propagation of combustion front in the presence of Magnetic field}

The preceding discussion is mainly a feasibility study for the possible
generation of the combustive phase transition front. 
Using it to be the initial condition, we now study the
evolution of the hydrodynamical combustion front in the 
presence of magnetic field. 
Following our recent work \cite{ritam1}, we choose a regime where the 
propagation front moves as a detonation front.
To examine such an evolution, we move to a 
reference frame in which the nuclear matter is at rest. The speed of the 
combustion front in such a frame is given by ${v}_{f}={-v}_{1}$ with $v_{1}$
being the velocity of the nuclear matter in the rest frame of the
front.

In this section we study the time evolution of the shock front within the 
special relativistic formalism in the presence of the static background 
magnetic field in the neutron star. The energy momentum tensor of an ideal 
fluid in the presence of a magnetic field can be written as the sum of the 
matter energy momentum tensor \(T_{\mu \nu}^{M}\) and the magnetic 
energy momentum tensor \(T_{\mu \nu}^{B}\). The matter energy momentum tensor 
can be written as:
\[ T_{\mu \nu}^{M}=\left(p+\epsilon\right)u_\mu u_\nu + p\eta_{\mu \nu} \]
where, the fluid four velocity is \(u_\mu = \Gamma (1,\vec{v})\),
\(\Gamma\) is the Lorentz factor and \(\vec{v}\) is the fluid three velocity.
The pressure and energy density \(p\) and \(\epsilon\) are evaluated in the 
rest frame of the fluid.\\
The magnetic energy momentum tensor can be written in terms of its components 
as:
\[ T_{00}=\frac{B^2}{8 \pi} \]
\[T_{0i}=0\]
\[T_{ij}=\frac{B^2}{8\pi}\delta_{ij}-\frac{B_{i}B_{j}}{4\pi}\]
In writing down the components of the magnetic tensor, we have assumed that the
electric field is zero in the nuclear matter rest frame. Hence \(T_{0i}\) which is 
simply the Poynting flux vector is zero.  
The equations of motion follow from the covariant conservation of the total
energy momentum tensor \(\partial_{\nu} T_{T}^{\mu \nu} = 0\). Thus we can 
write, \(\partial_{\nu}T_{\mu \nu}^{M}=-\partial_{\nu} T_{\mu \nu}^{B}.\) 
The expression on the right hand side in the above expression is simply the 
Lorentz four force \(f^{\mu}\). The zeroth component of the four force 
\(f^{0}\) is zero since there is no electric fields 
whereas the three-force is the Lorentz force due to magnetic field  
\( \vec{F}=(\nabla \times \vec{B})\times \vec{B}\). 

Using the above arguments we can write down the hydrodynamic equation 
for the propagation of the transition front \cite{ritam1} in the presence of 
the magnetic field:
 \begin{equation}
\frac{1}{\omega}(\frac{\partial\epsilon}{\partial\tau}
+v\frac{\partial\epsilon}{\partial r})+
\frac{1}{W^{2}}(\frac{\partial v}{\partial r}
+v\frac{\partial v}{\partial\tau})+2\frac{v}{r}=0
\label{7}\end{equation}

and 
\begin{equation}
\frac{1}{\omega}(\frac{\partial p}{\partial r}+
v\frac{\partial p}{\partial\tau})+
\frac{1}{W^{2}}(\frac{\partial v}{\partial\tau}+
v\frac{\partial v}{\partial r})=\frac{\langle F_r\rangle}{4 \pi \omega},
\label{8}\end{equation}
 where, $v=\frac{\partial r}{\partial\tau}$ is the front velocity
in the nuclear matter rest frame and $k=\frac{\partial p}{\partial\epsilon}$
is taken as the square of the effective sound speed in the medium. $p$, 
$\epsilon$, $\omega$ and $W$ are the pressure, energy density, enthalpy 
and Lorentz factor respectively. 

The effect of magnetic fields is included through the Lorentz 
force as given in eqn. \ref{8}. The energy equation is unaffected 
since we assume that the conductivity is formally infinite so that there 
are no dissipative effects. It is also to be noted that we have 
included the Lorentz force by considering its average over the polar angles. 
The Lorentz force term is not spherically symmetric. As a result, 
the shock speed would also vary with direction. As we are concerned
only with the radial propagation (spherically symmetric star), we have 
considered 
approximate angle averages to incorporate the effect of angle dependence. 
The details of the different configurations and the 
corresponding forms for the Lorentz force are detailed in the Appendix. 

Equations (\ref{7}) and (\ref{8}) can be rewritten as:
 
\begin{equation}
\frac{2v}{\omega}\frac{\partial\epsilon}{\partial r}
+\frac{1}{W^{2}}\frac{\partial v}{\partial r}(1+v^{2})+
\frac{2v}{r}=0
\label{9}\end{equation}

and 
\begin{equation}
\frac{n}{\omega}\frac{\partial\epsilon}{\partial r}(1+v^{2})
+\frac{2v}{W^{2}}\frac{\partial v}{\partial r}=\frac{\langle F_r\rangle}{4 \pi 
\omega}
\label{10}
\end{equation}

We finally get a single differential equation for the front velocity:

\begin{equation}
\frac{dv}{dr}=\frac{1}{W^2[4v^{2}-k(1+v^{2})^{2}]}\left[\frac{\langle F_{r} \rangle}
{4 \pi \omega}+\frac{k(1+v^{2})}{r}\right]
\label{11}
\end{equation}

The above equation is integrated with respect to $r$ from the center to the 
surface. The initial velocity is fixed through the jump conditions for 
different values of the central densities.  
The magnetic field is incorporated by specifying different configurations. 

\section{Results}

Equations (\ref{5})
and (\ref{6}) specify the respective flow velocities $v_{1}$
and $v_{2}$ of the nuclear and quark matter in the rest frame of
the front. This would give us the initial velocity of the front ($-v_{1}$),
at a radius infinitesimally close to the center of the star, 
in the nuclear matter rest frame.
The nuclear matter EOS (Walecka model) has been used to construct the 
static configuration of
compact star, for different central densities, by using the standard
Tolman-Oppenheimer-Volkoff (TOV) equations \cite{tol,open}. The velocity
at the center of the star should be zero from symmetry considerations.
On the other hand, the $1/r$ dependence of the $ \frac {dv}{dr} $,
in eq. (\ref{11})
suggests a steep rise in velocity near the center of the star. 
Equation (\ref{11}) is then integrated, with respect to $r$, starting
from the center towards the surface of the star.
The solution gives us the variation of the velocity with the position
as a function of time of arrival of the front, along the radius of
the star.

In this section, we present the results of the velocity profiles for different 
magnetic field strength for dipole configurations. We 
first consider a poloidal magnetic field which can be specified in terms of 
the field components as:
\begin{eqnarray}
B_r=\frac{B_0 \sin \theta}{r^3}   \nonumber \\
B_\theta= \frac{2 B_0 \cos \theta}{r^3}
\label{x}
\end{eqnarray}
In the magnetic field profile, $r$ is the distance from the center 
while \(B_0\) is related to 
the surface magnetic field \(B_{S}\) through \(B_0=B_{S} R^3\), where 
$R$ is the radius of the star. 
As the interior of the star is hidden from
direct observation we only know about the surface magnetic field of the 
star. Therefore, in this work we would always quote the surface magnetic 
field value. 
In our calculation the front starts from a distance of $100-200$m from the 
center of the star. Therefore surface field of $10^{10}$G and $10^{12}$G would have a 
value of $10^{16}$G and $10^{18}$G  respectively at the starting point of our calculation. These values are 
within the maximum allowed field value for the stability of magnetized neutron stars \cite{lai}. The 
angle averaged Lorentz force \(\langle F_r \rangle \) for this configuration
is \(\langle F_r \rangle = \frac{B_0^2}{12 \pi r^7}\) as evaluated in the 
appendix.

\begin{figure}
\vskip 0.2in
\centering
\includegraphics[width=3.0in]{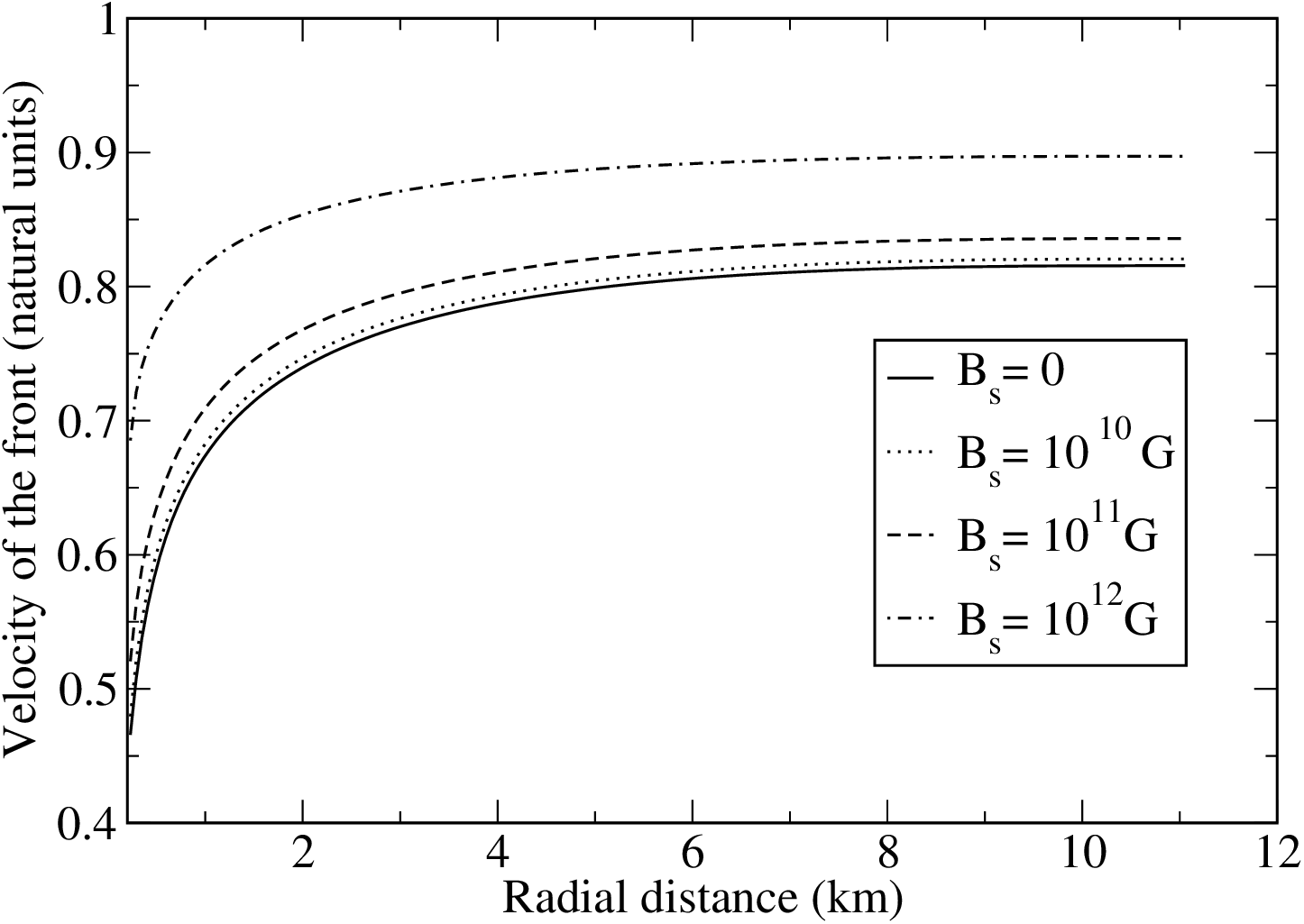}
\caption{Variation of velocity of the conversion front along the radial 
distance of the star for various values of the surface magnetic field.}
\label{diff-mag}
\end{figure}

\begin{figure}
\vskip 0.2in
\centering
\includegraphics[width=3.0in]{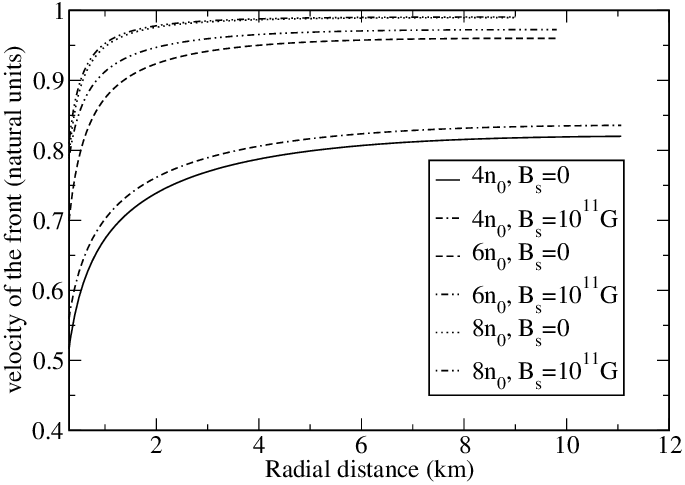}
\caption{Variation of velocity of the conversion front along the radial 
direction with a surface magnetic field of strength of $10^{11}$ Gauss for 
different values of the central density.}
\label{diff-den}
\end{figure}

\begin{figure}
\vskip 0.2in
\centering
\includegraphics[width=3.0in]{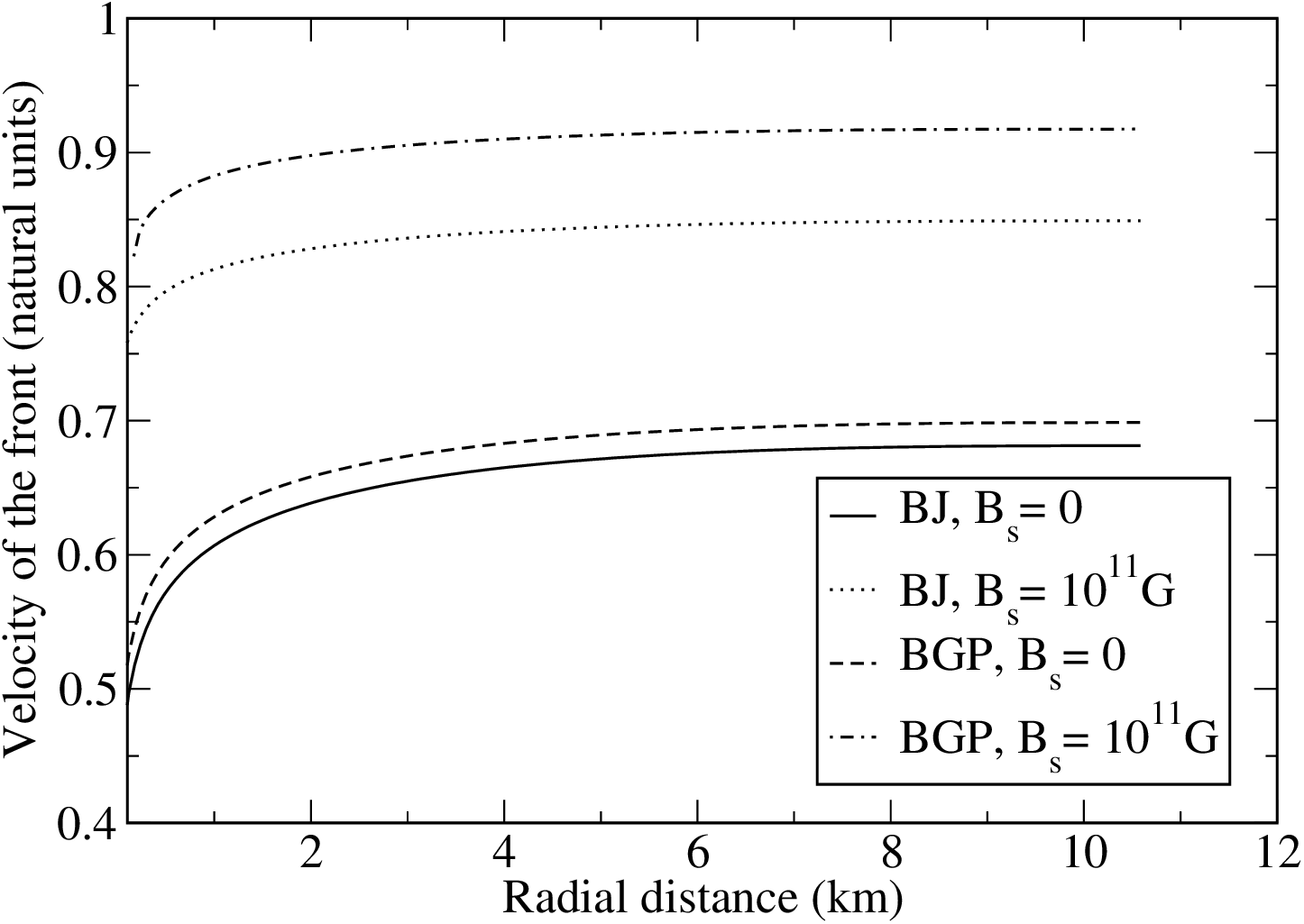}
\caption{Variation of velocity of the conversion front along the radial 
direction with a surface magnetic field of strength $10^{11}$ Gauss for two 
different EOS namely BGP and BJ.}
\label{diff-model}
\end{figure}

In fig. \ref{diff-mag}, we show the propagation 
velocity of the conversion front 
along the radius of the star with a central density of $4$ times 
nuclear matter saturation density. The solid curve denotes 
the SR calculation without the consideration of the magnetic field.
The dotted curve is for SR calculation with a surface magnetic field
of dipolar nature of strength $10^{10}$ Gauss. Below this field the effect of 
magnetic field on the conversion front is very negligible. The curve shows
that the magnetic field enhances the velocity of the front which is also
clear from equation (\ref{11}). Next we 
increase the magnetic field strength and the subsequent curves on the 
figure shows that the velocity of the conversion front increases with the 
increase of field strength. We also find that
the magnetic field effect is much pronounced at the center of the star.
This is because as we go towards the center the magnetic field increases 
as $r^3$, as given in eqn \ref{x}. So although the density increases the
magnetic field increases much rapidly towards the center of the star.   

Fig. \ref{diff-den} shows the velocity of the conversion front along the 
radial direction of the star for different central densities of the star. 
The solid
curve is for the central density $4$ times that of nuclear matter 
saturation density ($n_0$) without magnetic field effect. 
The dash-dash-dot 
curve is for the same central density with a surface magnetic field of 
$10^{11}$ Gauss. As expected we find that velocity of the front increases 
with the magnetic field. 
Next we increase the central density, and the two other set of curves are for 
the central densities corresponding to $6$ times and $8$ times that of 
nuclear matter saturation density respectively. The comparison with 
and without magnetic field is done with 
same value of the surface magnetic field ($10^{11}$ Gauss). We find that as 
the central density increases the velocity of the propagating front also 
increases. This is due to the fact that the velocity of the front  
depends on the value of $k$ (equation (\ref{11})), or the stiffness of the EOS.
For a high central density the star is more compact and therefore 
$k$ is much larger resulting in the further increase of the velocity.
We also find that for the same value of magnetic field as the density increases the effect of magnetic field decreases. This is because the denser the matter
greater the magnetic field has to be applied to have the same effect as the
matter pressure is higher at higher densities. So the magnetic pressure has
to be higher to be in comparison with the matter pressure to be able to show
significant change.

\begin{figure}
\centering
\includegraphics[width=3.0in]{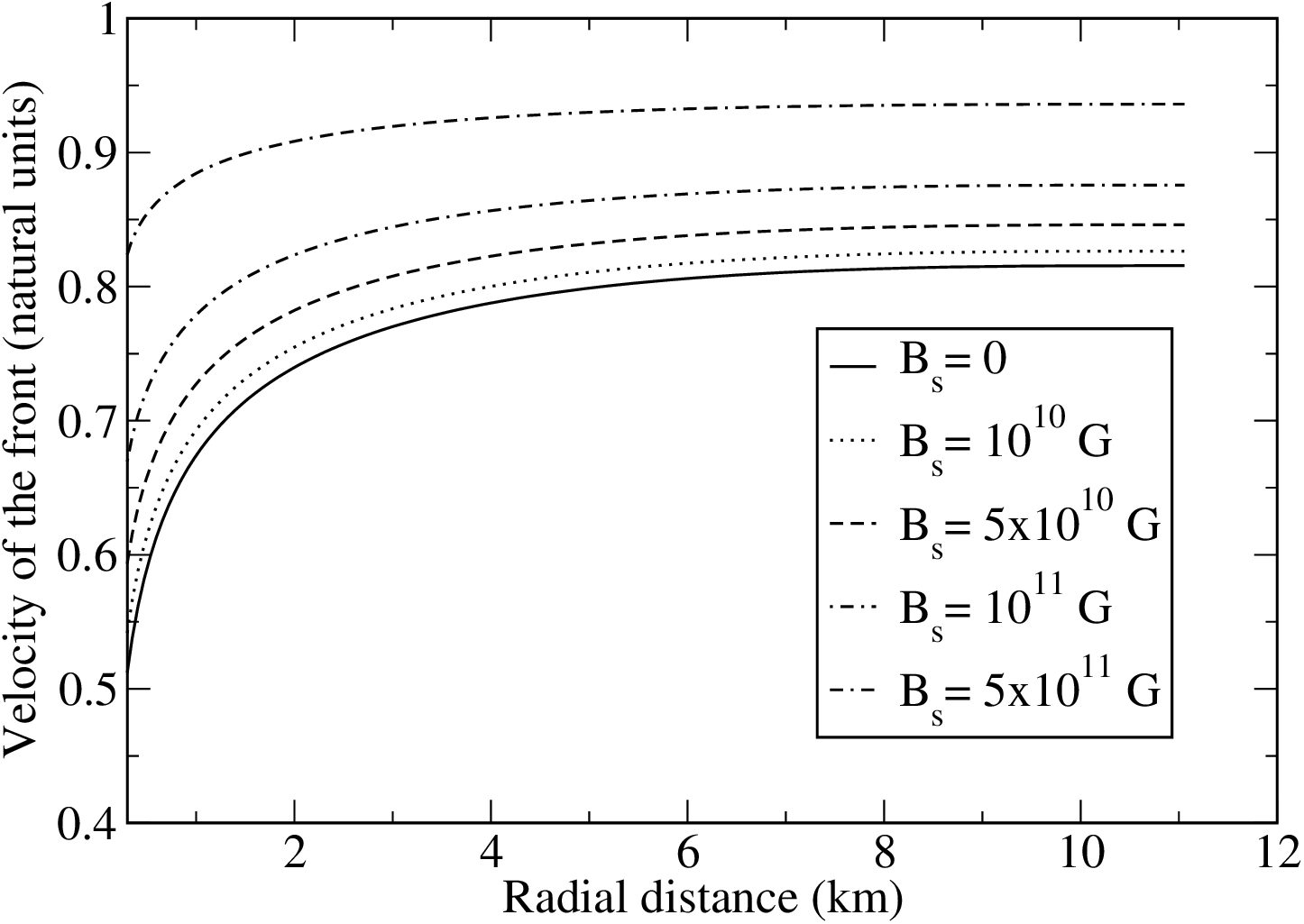}
\caption{Variation of velocity of the conversion front along the radial 
direction for different magnetic field strength for toroidal field.}
\label{tor}
\end{figure}

Similar calculation with central density $4$ times that of nuclear
saturation density is done for two other different EOS, Bowers, 
Gleeson and Pedigo (BGP) model \cite{bowers} and Bethe and Johnson
(BJ) model \cite{bethe}. Fig. \ref{diff-model} 
shows the velocity of the front for these 
two models with and without the effect of magnetic field. The surface
magnetic field applied for the two cases is same ($10^{11}$ Gauss). 
We again find that the magnetic field increases the velocity of 
the propagation front. As the BGP model being much 
stiffer than the BJ model EOS, the velocity of the front
is larger for the BGP model compared to that of the BJ model. 
Therefore we find that 
the velocity of the front depends on the central density as well as the 
EOS we choose. The velocity increases with the increase in magnetic field.
Below a certain value of the field the effect on the conversion front is 
not at all relevant and for a very high value of magnetic field (which may 
be found in magnetars) the velocity of the front attains a velocity close 
to that of the speed of light.

Little is known about the magnetic field structure in neutron star, and 
therefore it is not good enough to assume that only poloidal field would 
be present. In fact, a proto-neutron star dynamo \cite{thompson} is unlikely 
to generate purely poloidal 
field and differential rotation will easily wrap any poloidal field and 
generate strong toroidal field \cite{wheeler}. Therefore it seems realistic 
to consider effect of magnetic field configuration consisting of poloidal 
and toroidal components.
The toroidal magnetic field grows from the poloidal dipole 
field. This can be possible for the differential rotation 
($\omega_d$) usually present in  the 
neutron stars \cite{rasio,shibata}. The velocity due to the 
differential rotation can be written as 
\begin{eqnarray}
V_d=\omega_d r\hat{\phi}.
\end{eqnarray}
As nothing specific is known about the differential rotation, 
for simplification, we assume $\omega_d$ to be constant and equal to $1$. 
So that tt can be taken care of along with the normalization constant.
The toroidal field $B_T$ due to this differential rotation 
is given by 
\begin{eqnarray}
B_T=\nabla \times(V_d \times B)
\end{eqnarray}
where $B$ is the dipole field. Hence we get, 
\begin{eqnarray}
B_T=-\frac{B_0sin\theta}{r^3}\hat{\phi}
\end{eqnarray}
with some constant which we have normalized. Using the equation given in 
the appendix for the toroidal field, the 
angle averaged Lorentz force \(\langle F_r \rangle \) for this configuration
is found to be \(\langle F_r \rangle = \frac{B_0^2}{2 \pi r^7}\). 
In fig. \ref{tor} 
we have plotted the velocity of the conversion front for the toroidal field.
As we have averaged out the Lorentz force over the polar angle the angle 
averaged Toroidal field comes out to be of the same form as the poloidal 
force (see appendix). 
The nature of the curve remains same, as the angle averaged toroidal field 
is just six times than that of the poloidal field. Therefore a much lesser 
value of surface magnetic field is needed to generate the same effect in the 
conversion front velocity than that needed for the poloidal field. However 
the main factor for the toroidal field configuration remains the time and 
the value of the differential rotation of the star.   

\section{Summary and Discussion}

To summarize, we have studied the effect of magnetic field in the phase 
transition of NS to two-flavour QS. We have incorporated a Lorentz force in 
the equation of motion for the propagation of a front across the star 
converting the NS to two-flavour QS. We have assumed a field of dipolar 
nature with a field strength of maximum $10^{18}$G near the center of 
the star from where our calculation begins.

We find that the velocity of the front increases with the increase of 
magnetic field strength. We have also calculated the velocity of the 
front for different central density with same magnetic field of 
$10^{11}$G at the surface. We find that the velocity of the front 
increases with the increase in central core density of the star. We have 
done a analytical calculation for different 
types of field configuration (in the appendix) and for a initial field
of dipolar nature, and have shown our results for poloidal and toroidal 
configuration.  For a field of dipolar nature
there is an increment of about $10$ percent in the front velocity for a field
strength of $10^{18}$G near the center. In our calculation of angular 
averaging we find that the toroidal angle average field is $6$ times more 
stronger that the poloidal angle average value. Therefore the effect of 
magnetic field for toroidal case is much more pronounced.

It should be mentioned at this point that an 
earlier set of authors \cite{lugones} studied the effect of an idealized
configuration of the magnetic field - a constant magnetic field of moderate
strength of $\sim 10^{12-13}G$ in a dipolar geometry - on the Rayleigh-Taylor 
(RT) instability. Through an order of magnitude estimate, they concluded that
there would be a difference in the velocity of the combustion front along
polar and equatorial directions. 
Our calculation is based on initial dipolar magnetic field with a 
high value of magnetic field at the center. 
Our previous work with magnetic field was interesting but the origin of
magnetic force on the equation of motion was truly put by hand.
As this is a first attempt of such kind, to have an overall understanding of 
the magnetic effect on propagation front, starting from the tensorial 
conservation equation, we have done averaging in the 
angular direction. For small 
value of magnetic field it is quite acceptable, but for high value of 
magnetic field  effect of angular dependence may become more pronounced. 
The present study 
gives a good estimation of the effect of magnetic field on the propagation
speed through an order of magnitude estimation. To get finer details, we 
need to solve general MHD equation in presence of all the components.
As the exact nature of the field configuration is not known,
we need to carry out similar calculation with other  
forms of magnetic field. 
Furthermore, such high magnetic fields can even modify the 
Einstein equation for the metric as the matter is then no longer an ideal 
fluid. All such a calculation is on our immediate agenda.

\subsection{Appendix : Angle -averaged radial Lorentz force}

The Lorentz force due to a magnetic field B is given by 
\begin{equation}
{\vec{F}}=\frac{(\nabla\times \vec{B})\times\vec{B}}{4 \pi}
\label{eq:Lorentz}
\end{equation}
This can also be written as :
\begin{equation}
4 \pi \vec{F} = -\frac{1}{2} \nabla (B^2)+(\vec{B}.\nabla)\vec{B}
\label{eq:Lorentz1}
\end{equation}

Using spherical coordinates, the radial component of the Lorentz force can be 
simplified as \cite{gr2010},
\begin{equation}
4 \pi F_{r}=\nabla.(B_{r} \vec{B})-
\frac{{B_{T}}^2}{r}-\frac{1}{2}\frac{\partial}{\partial r}(B^2)
\label{eq:LorentzRadial1}
\end{equation}
Here, ${B_{T }}^2={B_{\phi}}^2+{B_{\theta}}^2$, denotes the transverse 
part of the expression. The angle averaged Lorentz force is defined as: 
\begin{equation}
\langle F_{r}\rangle=\int \frac{d \Omega}{4\pi} F_{r}
\label{eq:angleaveF_r}
\end{equation}
Taking the angular average of Eq~({\ref{eq:LorentzRadial1}}), we get:
\begin{equation}
4 \pi \langle F_{r} \rangle = \langle \nabla.(B_{r} \vec{B}) \rangle - \frac{\langle {B_{T}}^2 
\rangle}{r}
-\frac{1}{2}\frac{\partial}{\partial r}(\langle B^2 \rangle)
\label{eq:AngAveFr}
\end{equation}
The angle average of 
$\langle \nabla.(\vec{B}B_{r})\rangle$ can be easily evaluated. Let
\begin{equation}
f(r)=\langle \nabla.(\vec{B}B_{r})\rangle =\int\frac{d \Omega}{4\pi}\nabla.(\vec{B}B_{r})
\label{eq:fr}
\end{equation}
Multiplying both sides by $r^{2} dr$ and integrating,
\begin{equation}
\int dr r^{2} f(r)=\int \frac{dV}{4\pi} \nabla.(\vec{B}B_{r}).
\label{eq:frvol}
\end{equation}
Using Stokes theorem, the right hand side of the volume integral can be 
expressed as a surface integral,
\begin{equation}
\int dr r^{2} f(r)=r^{2} \int \frac{d\Omega}{4\pi} \hat{r}.(\vec{B}B_{r})
\label{stokes}
\end{equation}
Now, differentiating both sides w.r.t $r$ we get:
\begin{equation}
f(r)=\frac{1}{r^{2}} \frac{d}{dr}(r^{2}\langle B_{r}^{2}\rangle )
\label{eq:StokesDiff}
\end{equation}
Using Eq~(\ref{eq:angleaveF_r}), Eq (18) and the fact that 
$B^2=B_r^2+B_\theta^2+B_\phi^2$, the final expression 
for the angle-averaged radial force can be written in terms of the angle average 
of the different components as: 
\begin{equation}
4 \pi \langle F_{r}\rangle = \frac{1}{2} \frac{d}{dr}\left[\langle B^2 \rangle - 2 \langle 
{B_{T}}^2 \rangle\right]
+\frac{2 \langle B^2 \rangle - 3\langle {B_{T}}^2 \rangle}{r}
\label{eq:Frave}
\end{equation}
 
The above expression for the angle-averaged radial Lorentz force can be simplified 
for different magnetic field configurations some of which we enumerate below:
\begin{enumerate}
\item $\langle {B_{\theta}}^{2}\rangle = \langle {B_{\phi}}^{2}\rangle \simeq 0$ (nearly radial field):\\
\indent In this case, $\langle B^{2}\rangle \simeq \langle {B_{r}}^{2} \rangle$. Hence
\begin{equation}
\langle F_{r}\rangle = -\frac{d}{dr}\left\langle \frac{B^{2}}{8 \pi}\right\rangle - 
\frac{4}{r}\left\langle \frac{B^{2}}{8 \pi} \right\rangle
\label{eq:Frave1}
\end{equation}
\item $\langle {B_{r}}^{2}\rangle=\langle {B_{\theta}}^{2}\rangle =\langle 
{B_{\phi}}^{2}\rangle=\frac{1}{3}\langle B^{2}\rangle $:\\
\indent In this case, the simplified expression is,
\begin{equation}
\langle F_{r} \rangle = -\frac{1}{3}\frac{d}{dr}\left\langle \frac{B^{2}}{8 
\pi}\right\rangle
\end{equation}
\item $ \langle {B_{\phi}}^{2} \rangle = 0 $(poloidal field):\\
\indent In this case, the simplified expression is,
\begin{equation}
\langle F_{r} \rangle = \frac{1}{8 \pi}\frac{d}{dr}[B_r^2-B_{\theta}^2] + 
\frac{2 B_r^2 - B_{\theta}^2}{4 \pi r}
\end{equation}
For the special case of a dipole magnetic field, which we assume for calculations 
in this paper, we have 
\[\langle {B_{r}^2} \rangle =\frac{B_0^2}{r^6}\int \frac{d \Omega}{4\pi}\sin^2 \theta=\frac{2B_0^2}{3r^6}\]
\[\langle {B_{\theta}^2} \rangle=\frac{4 B_0^2}{r^6}\int\frac{d \Omega}{4 \pi}\cos^{2} \theta=\frac{4B_0^2}{3r^6}\] 
The Lorentz force thus can be expressed as :\[\langle F_{r} \rangle = \frac{B_0^2}{12 \pi r^7}\]

\item $\langle {B_{r}}^2\rangle = \langle {B_{\theta}}^2 \rangle = 0$ 
(toroidal field) :\\
\indent In this case, the expression simplifies to:
\begin{equation}
\langle F_{r} \rangle = -\frac{d}{dr}\left\langle\frac{B_{\phi}^2}{8 
\pi}\right\rangle - \frac{B_{\phi}^2}{4 \pi r}
\end{equation}
For the special case of dipole field, we get
\[\langle F_{r} \rangle = \frac{B_0^2}{2 \pi r^7}\]
\end{enumerate}

\subsection{Acknowledgments}
R.M. thanks Grant No. SR/S2HEP12/2007, funded by DST, India for 
financial support. 
S.K.G. and S.R. thank Department of Science \& Technology, Govt. of India for 
support under the IRHPA scheme.

\end{document}